\documentclass[prb,onecolumn,superscriptaddress,showpacs,unsortedaddress]{revtex4}
\usepackage{mathtools}
\usepackage{amsmath}
\usepackage{graphicx}
\usepackage{float}

\DeclareMathOperator{\erf}{erf}

\begin{document}

\title{Extended formalism for simulating compound refractive lens-based x-ray microscopes}

\author{Hugh Simons}
\author{Sonja Rosenlund Ahl}
\author{Henning Friis Poulsen}
\affiliation{Department of Physics, Technical University of Denmark, Lyngby 2800 kgs, Denmark}
\author{Carsten Detlefs}
\affiliation{European Synchrotron Radiation Facility, 71 Avenue des Martyrs, Grenoble 38000, France}

\date{\today}

\begin{abstract}
We present a comprehensive formalism for the simulation and optimisation of CRLs in both condensing and full-field imaging configurations. The approach extends ray transfer matrix analysis to account for x-ray attenuation by the lens material. Closed analytical expressions for critical imaging parameters such as numerical aperture, vignetting, chromatic aberration and focal length are provided for both thin- and thick-lens imaging geometries.
\end{abstract}

\maketitle

\section{Introduction}
Compound refractive lenses (CRLs) are predominantly used as micro- and nano-focusing lenses in scanning beam microscopy \cite{Ice2011}. However, there is a growing use of CRLs as imaging objectives in hard x-ray microscopes\cite{Simons15}. Improving the design of CRLs and their implementation in x-ray microscopes requires a thorough understanding of the optical principles of CRLs as well as convenient and simple analytical expressions for predicting and optimizing imaging performance across a wide variety of optical configurations.

Different approaches have addressed the optical theory of CRLs and CRL-based imaging systems, such as ray-transfer matrices (RTMs) \cite{Protopopov98} (including with Gaussian beams \cite{Poulsen2014}), Monte Carlo ray tracing \cite{SanchezdelRio2012}, wavefront propagation \cite{Kohn2003} and others \cite{Lengeler99}. While each have merits, no single formalism provides the necessary combination of closed analytical expressions, applicability to both condensing and full-field imaging systems, and consideration of the full range of possible geometries (i.e. both the thin and thick-lens imaging conditions).

This work presents a step-by-step derivation of a formalism for CRL-based imaging systems. Intended as a tutorial for beginners in the field, it utilises an RTM approach to model typical x-ray imaging systems in a manner that accounts for both thin-lens and general (i.e. thick-lens) cases. We first introduce the RTM formalism, then derive the the ray path within the CRL. Following this, the CRL formalism is placed in the context of a complete imaging system, and relevant expressions for acceptance functions, resolution and aberration are provided. These expressions are key requirements for efficient parametric optimisation of the CRL and imaging geometry, which could ultimately provide suggestions for future lens development routes. 

\section{The RTM approach}
This RTM approach assumes a 1D geometry, valid for axisymmetric and planar CRLs with parabolic curvatures (Fig. 1). The CRL is comprised of $N$ parabolic lenses arranged in a linear array, where each lens has a radius of curvature $R$, aperture $2Y$ and centre-to-centre distance between successive lenslets $T$ so that $Y = \sqrt{RT}$. Lenslets may also have a gap between their apices, given as $T_\mathrm{web}$ and a physical lenslet thickness $T_\mathrm{phys} < T$. The physical aperture of the lenslet $2Y_\mathrm{phys}$ is therefore given by $Y_\mathrm{phys} = \sqrt{R(T_\mathrm{phys}-T_\mathrm{web})}$. 

\begin{figure}[H] \label{fig1}
\centering\includegraphics[width=10cm]{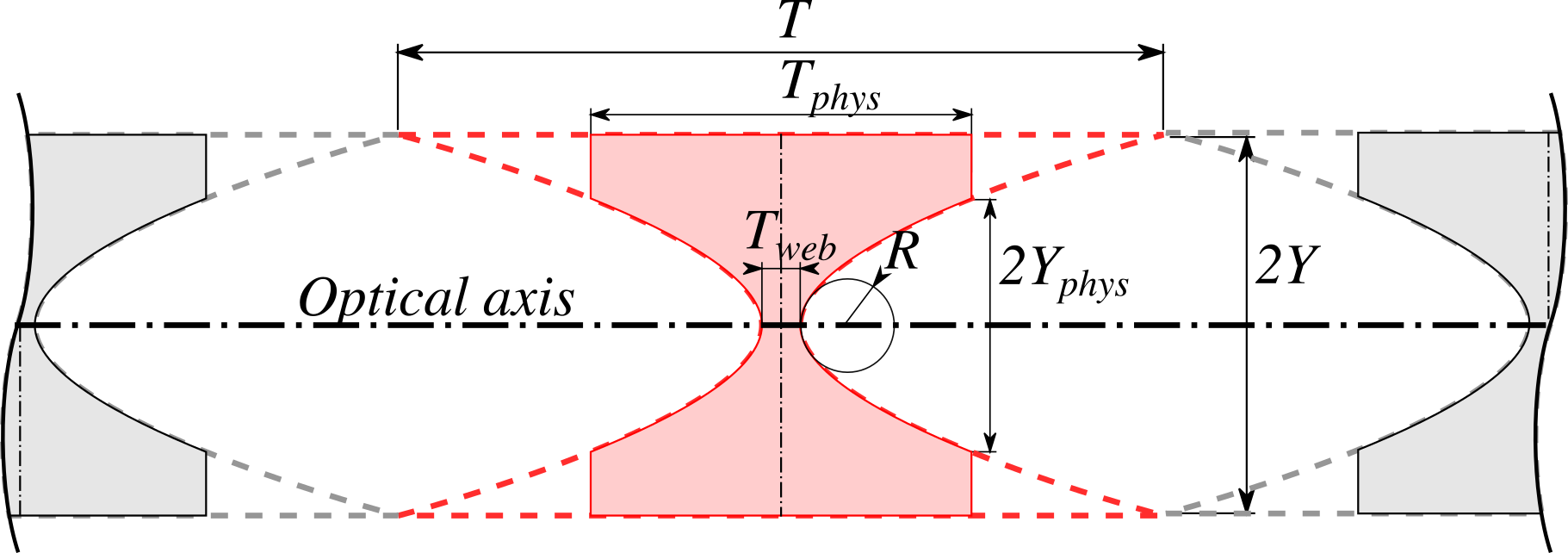}
\caption{Geometry of the 1D radially symmetric (i.e. axisymmetric) refractive lenslet. A single refracting lenslet element is shown in blue, annotated with symbolic dimensions.}
\end{figure}

Photons are treated as rays with position $y$ and angle $\alpha$. The RTM then describes an optical system as a matrix $ \mathbf{M}$ that transforms an incident ray $(y_0,\alpha_0)$ into an exit ray $(y_1,\alpha_1)$. 

\begin{equation} \label{eq1}
	\begin{bmatrix}
		y_1 \\
		\alpha_1
	\end{bmatrix}
	= \mathbf{M}
	\begin{bmatrix}
		y_0 \\
		\alpha_0
	\end{bmatrix}
\end{equation}

RTM analysis inherently assumes the following:
\begin{enumerate}
\item{Paraxial rays, i.e. that the angle between rays and the optical axis is small such that $\sin{(\alpha)} \approx \alpha$ and $\cos{(\alpha)} \approx 1$. This is justified by the weak refractive decrement $\delta$ and consequently small numerical aperture of x-ray CRLs.}
\item{Thin lenslets, i.e. that the focal length $f$ of each lenslet is much larger than $T$. This is valid for the vast majority of lenslet geometries and the full x-ray energy regime relevant to CRLs (i.e. $E > 15$ keV)}
\item{Reflection and diffraction are negligible. This is well-justified in other approaches e.g. Ref. \cite{Schroer2005b}}
\end{enumerate}

We first derive the ray transfer matrix of a single, thin lenslet. Its focal length $f$ is given in terms of the lenslet radius $R$ and the refractive decrement $\delta$ as:
\begin{equation} \label{eq2}
	f = \frac{R}{2 \delta} 
\end{equation}

The corresponding transfer matrix $\mathbf{M}$ can be described as the combination of matrices corresponding to a free space propagation by half the distance between lenslet centres, $T/2$, followed by focusing with focal length $f$ and another free space propagation by $T/2$:
\begin{equation} \label{eq3}
	\mathbf{M} = 
	\begin{bmatrix}
		1 & T/2\\
		0 & 1
	\end{bmatrix}
	\begin{bmatrix}
		1 & 0\\
		-1/f & 1
	\end{bmatrix}
	\begin{bmatrix}
		1 & T/2\\
		0 & 1
	\end{bmatrix}
	=
	\begin{bmatrix}
		1-\frac{T}{2f} & f(1-(1-\frac{T}{2f})^2)\\
		\frac{-1}{f} & 1-\frac{T}{2f}
	\end{bmatrix} 
\end{equation}

The transfer matrix of the CRL, $\mathbf{M}^N$ is the combined behaviour of $N$ such lenslets. This can be calculated through the matrix eigendecomposition theorem, where $\mathbf{P}$ is a matrix comprising the eigen\emph{vectors} of $\mathbf{M}$ and $\mathbf{D}$ is a diagonal matrix comprising the eigen\emph{values} of $\mathbf{M}$.
\begin{equation} \label{eq4}
	\mathbf{M}^N = \mathbf{M} \mathbf{M} ... \mathbf{M}
	= (\mathbf{M})^N
	= \mathbf{P} \mathbf{D}^N \mathbf{P^{-1}} 
\end{equation}

The eigenvalues $E_{\pm}$ - which are a complex conjugate pair if $4f > T$ - may then be expressed as:
\begin{equation} \label{eq5}
	E_\pm = 1 - \frac{T}{2f} \pm i \sqrt{1-\left( 1-\frac{T}{2f}\right)^2}
	= \cos(\varphi)\pm i \sin(\varphi)
	= \exp{(\pm i \varphi)}
\end{equation}

where $\varphi$ is the phase angle of the complex eigenvalues, which may be expressed as:
\begin{equation} \label{eq6}
	\varphi = \tan^{-1}{\left(\frac{\sqrt{1-\left( 1-\frac{T}{2f}\right)^2}}{1 - \frac{T}{2f}}\right)}
	\approx	\sqrt{\frac{T}{f}}
\end{equation}

The corresponding eigenvectors $\mathbf{r}_{\pm} = (y_{\pm} , \alpha_{\pm} ) $ are given by $y_{\pm} = \mp i \alpha_{\pm} f \sin \varphi$. From Eq. \ref{eq5} the expression for $\mathbf{M}^N$ then becomes
\begin{equation} \label{eq7}
	\begin{split}
		\mathbf{M^N} &= 
		\begin{bmatrix}
			1 & 1 \\
			\frac{1}{if\sin{\varphi}} & \frac{-1}{if\sin{\varphi}}
		\end{bmatrix}
		\begin{bmatrix}
			\exp{(-iN\varphi)} & 0 \\
			0 & \exp{(iN\varphi)}
		\end{bmatrix}
		\begin{bmatrix}
			1 & 1 \\
			\frac{1}{if\sin{\varphi}} & \frac{-1}{if\sin{\varphi}}
		\end{bmatrix}^{-1}\\
		&=
		\begin{bmatrix}
			\cos{(N \varphi)} & f \sin{\varphi} \sin{(N \varphi)} \\
			\frac{-\sin{(N \varphi)}}{f \sin{\varphi}} & \cos{(N \varphi)}
		\end{bmatrix} 
	\end{split}
\end{equation}

Note also that the basic approximation of the RTM approach implies that
\begin{equation} \label{eq8}
\begin{split}
	 \sin{\varphi} &= \varphi = \sqrt{\frac{T}{f}} = \frac{1}{\sqrt{2\delta}} \frac{Y}{f} \\	
	 f \sin{\varphi} &= f \varphi  = \sqrt{T f} = \frac{Y}{\sqrt{2\delta}}
\end{split}
\end{equation}

Here, $\sqrt{2\delta}$ is the critical angle for total external reflection and the numerical aperture for an ideal elliptical lens \cite{AlsNielsen2010} corresponding to $R=T$ in the parabolic approximation. If there is no space between lenslets, $Y/f$ is half the numerical aperture of one lenslet and $1/(f\varphi)$ is the refractive power of the CRL per unit length \cite{Lengeler99}. 

Hence, from Eqs. \ref{eq7} and \ref{eq8}, $\mathbf{M}^N$ can then be expressed as:
\begin{equation} \label{eq9}
	\mathbf{M}^N =
	\begin{bmatrix}
		\cos{(N \varphi)} & f \varphi \sin{(N \varphi)} \\
		\frac{-\sin{(N \varphi)}}{f \varphi} & \cos{(N \varphi)}
	\end{bmatrix}
\end{equation}

\section{Derivation of the ray path within CRLs}
To calculate the position of a ray at the centre of the  $n^{th}$ lenslet, we determine the RTM of the CRL after the lenslet and then back-propagate by half a lenslet distance (i.e. $-T/2$). The ray vector $(y_n,\alpha_n)$ as a function of the incident ray $(y_0,\alpha_0)$ is then: 
\begin{equation} \label{eq10}
	\begin{bmatrix}
		y_n \\
		\alpha_n
	\end{bmatrix}
	=
	\begin{bmatrix}
		1 & -T/2\\
		0 & 1
	\end{bmatrix}
	\mathbf{M}^n
	\begin{bmatrix}
		y_0 \\
		\alpha_0
	\end{bmatrix}
	= 
	\begin{bmatrix}
		M_{11}^n -\frac{T}{2} M_{21}^n & M_{12}^n -\frac{T}{2} M_{22}^n \\
		M_{21}^n & M_{22}^n
	\end{bmatrix}
	\begin{bmatrix}
		y_0 \\
		\alpha_0
	\end{bmatrix}
\end{equation}

Meaning that the lateral position $y_n$ is given by:
\begin{equation} \label{eq11}
	\begin{split}
		y_n 
		&=  y_0 \left(M_{11}^n -\frac{T}{2} M_{21}^n\right) + \alpha_0 \left( M_{12}^n -\frac{T}{2} M_{22}^n \right) \\
		&= y_0\left(\cos(n \varphi) + \frac{T \sin(n \varphi)}{2f\varphi} \right) + \alpha_0\left(f\varphi \sin(n \varphi) - \frac{T}{2} \cos(n \varphi) \right) \\
		&= y_0\cos{\left(\left(n-\frac{1}{2}\right)\varphi\right)} + \alpha_0 f \varphi \sin{\left(\left(n-\frac{1}{2}\right)\varphi\right)}
	\end{split}
\end{equation}

Since:
\begin{equation} \label{eq12}
\begin{split}
 \cos(n \varphi) + \frac{T \sin(n \varphi)}{2f\varphi}  & = \cos(n\varphi) + \frac{\varphi}{2} \sin(n \varphi) = \cos \left(\left( n- \frac12 \right) \varphi\right) \\
f\varphi \sin(n \varphi) - \frac{T}{2} \cos(n \varphi)  &=  f \varphi \left(\sin(n\varphi) - \frac{\varphi}{2} \cos(n \varphi) \right)  = f \varphi \sin \left(\left( n- \frac12 \right) \varphi\right) 
\end{split}
\end{equation}

Trigonometric identities (see Appendix) then give:
\begin{equation} \label{eq13}
	y_n = \sqrt{ y_0^2 + \left(  \alpha_0 f \varphi \right) ^2 } \sin \left(  \left(n -\frac{1}{2} \right)  \varphi + \tan^{-1} \left( \frac{y_0 }{\alpha_0 f \varphi}  \right)       \right)
\end{equation}

All rays through the CRL will therefore have a sinusoidal trajectory with a period of $\frac{2\pi Y}{\sqrt{2\delta}}$ that is proportional to the aperture Y. Likewise, for $\alpha_n$:
\begin{equation} \label{eq14}
	\alpha_n = \frac{1}{f\varphi} \sqrt{y_0^2+\left(\alpha_0f\varphi\right)^2} \sin{\left(n\varphi - \tan^{-1}{\left(\frac{y_0}{\alpha_0 f \varphi}\right)}\right)} \\
\end{equation}

If at any point at the trajectory, $\mid y_n \mid >  Y_\mathrm{phys}$, the ray will not be refracted, but either absorbed or contribute to the general background after the lens. From this, the criterion for any ray to participate in the focusing/imaging process becomes:
\begin{equation} \label{eq15}
			\mid y \mid_{max}  \le  Y_\mathrm{phys}
\end{equation}

As we will show, the imaging condition implies that there at most can be one extremum in practical systems.  Hence, the maximum value of $y$, $y_{max}$ is:  
\begin{equation} \label{eq16}
	\mid y \mid_{max} = \max \left( \mid y_0 \mid, \mid  y_N \mid ,  \sqrt{ y_0^2 + \left(  \alpha_0 f \varphi \right) ^2 }      \right)  \\
\end{equation}

\section{Derivation of focal length}
The focal length is the distance from the lens exit $f_N$ at which a refracted ray with initial angle $\alpha_0=0$ will intersect with the optical axis at $y=0$. This can be conveniently extracted from $\mathbf{M}^N$ \cite{Gerrard2012}:
\begin{equation} \label{eq17}
	f_N = -\frac{M^N_{11}}{M^N_{21}}
	= f \varphi \cot{(N \varphi)}
\end{equation}

Notably, $f_N$ is measured from the exit point of the CRL along the optical axis, $NT/2$ from the CRL centre. For small $N \varphi$ the expression above  approximates to $f_N = \frac{R}{2\delta N} - \frac{N T}{3}$, for which negligible total lens thickness $NT$ reproduces equivalent expressions derived by alternative means, e.g. from \cite{Lengeler99}.  

\section{Derivation of the attenuation profile}
For a single lenslet, the attenuation $U$ of a ray passing through a lenslet with linear absorption coefficient $\mu$ and local thickness $t$ at perpendicular distance $y$ from the optical axis can be described by the Beer-Lambert law and a box function $H$, which fully attenuates any ray that exceeds the physical aperture $Y_\mathrm{phys}$:

\begin{equation} \label{eq18}
	U(y) = \exp{[-\mu t(y)]} \times H{\left(\frac{y}{Y_\mathrm{phys}}\right)},
\end{equation}

where the box function $H$ is:
\begin{equation} \label{eq19}
	H(x) = \left\{
        \begin{array}{ll}
            1 & \text{for} \vert x \vert \leq 1 \\
            0 & \text{for} \vert x \vert > 1
        \end{array}
    \right.
\end{equation}
 
The paraxial assumption is that the rays propagate each lenslet at a constant distance to the optical axis. The parabolic profile of the lenslets then gives the thickness function $t(y)$:
\begin{equation} \label{eq20}
	t(y) = T_\mathrm{web} + \frac{y^2}{R}  \hspace{0.5cm} \text{for} \hspace{0.5cm} \mid y  \mid < Y_\mathrm{phys}
\end{equation}

Thus, neglecting the effects of air between lenslets (i.e. assuming high energy x-rays), a single lenslet will attenuate the ray according to:
\begin{equation} \label{eq21}
	U(y) = \exp{(-\mu T_\mathrm{web})} \times \exp{\left(\frac{-\mu y^2}{R}\right)} \times H{\left(\frac{y}{Y_\mathrm{phys}}\right)}
\end{equation}

The cumulative attenuation $U_N(y)$ of a ray as it travels through $N$ lenslets is then the product of the individual attenuation contributions from each lenslet:
\begin{equation} \label{eq22}
	U_N = \prod_{n=1}^N U(y_n)
	= \exp{(-N \mu T_\mathrm{web})} \times \exp{\left(\frac{-\mu}{R} \sum_{n=1}^N y_{n}^2\right)} \times \prod_{n=1}^N H{\left(\frac{y_n}{Y_\mathrm{phys}}\right)},
\end{equation}

where $y_{n}$ is the position of the ray at the centre of the $n^{th}$ lenslet. The central expression for $\frac{\mu}{R} \sum_{n=1}^N y_n^2$ is a quadratic term, which can be expressed as:
\begin{equation} \label{eq23}
	\frac{\mu}{R} \sum_{n=1}^N y_n^2 = A_N   \alpha_0^2 +  B_N  \alpha_0 y_0 + C_N y_0^2
\end{equation}

The coefficients $A_N$, $B_N$ and $C_N$ are geometric sums which, since $1+\varphi^2 \approx 1$, have solutions to second order in $\varphi$ (see Appendix for identities):
\begin{equation} \label{eq24}
\begin{split}
	A_N 
	&= \frac{\mu}{R} \sum_{n=1}^N( M_{12}^n -\frac{T}{2} M_{22}^n )^2 \\
	&= \frac{N\mu(f\varphi)^2}{2R} \left( 1 -  \text{sinc}( 2N\varphi) \right) \\ 
	&= \frac{N\mu T}{4\delta} \left( 1 -  \text{sinc} ( 2N\varphi) \right)		
\end{split}
\end{equation}

\begin{equation} \label{eq25}
\begin{split}
	B_N  
	&= \frac{\mu}{R} \sum_{n=1}^N  (M_{11}^n -\frac{T}{2} M_{21}^n)( M_{12}^n -\frac{T}{2} M_{22}^n ) \\
	&= \frac{\mu }{2\delta} \left(1-\cos(2N \varphi)\right)	
\end{split}
\end{equation}

\begin{equation} \label{eq26}
\begin{split}
	C_N 
	&=   \frac{\mu}{R} \sum_{n=1}^N  (M_{11}^n -\frac{T}{2} M_{21}^n)^2 \\
	&=  \frac{N\mu}{2R} \left( 1 +  \text{sinc}(2N\varphi) \right)			
\end{split}
\end{equation}

Hence, the attenuation function can be expressed as a 2D tilted Gaussian function with clipped' boundaries defined by $y_n/Y_\mathrm{phys}$:
\begin{equation} \label{eq27}
	U_N = \exp{(-N \mu T_\mathrm{web})} \times \exp{\left(A_N \alpha_0^2 + B_N \alpha_0 y_0 + C_N y_0^2\right)} \times \prod_{n=1}^N H{\left(\frac{y_n}{Y_\mathrm{phys}}\right)},
\end{equation}

\section{Derivation of physical and effective apertures}
The physical aperture of the CRL corresponds to the spatial acceptance (i.e. in $y$) for a parallel incident beam (i.e. $\alpha_0 = 0$). When $Y_\mathrm{phys}$ is sufficiently large, the spatial acceptance has a Gaussian profile in $y$:
\begin{equation} \label{eq28}
	U_\mathrm{phys} = \exp{(-\mu N T)}\exp{\left(\frac{-y_0^2}{2\sigma_D^2}\right)}
\end{equation}

In such (common) cases, the physical aperture is given by the RMS $\sigma_D$:
\begin{equation} \label{eq29}
	\sigma_D =  \frac{1}{\sqrt{2 C_N}} = \sqrt{\frac{R}{\mu N}}\frac{1}{\sqrt{1 + \mathrm{sinc} (2N\varphi)}}\\
\end{equation}

The effective aperture is the diameter $D_\mathrm{eff}$ of a cylindrical pupil function that has the same total transmission as the CRL \cite{Lengeler99}. Its expression is found by solving:
\begin{equation} \label{eq30}
	\pi\left(\frac{D_\mathrm{eff}}{2}\right)^2 = 2\pi \exp{(-\mu N T)} \int_{0}^{Y_\mathrm{phys}}{\exp{\left(\frac{-y_0^2}{2\sigma_D^2}\right)} y_0 d y_0}
\end{equation}

Hence:
\begin{equation} \label{eq31}
	D_\mathrm{eff} = 2\sqrt{2}\sigma_D \exp{\left(\frac{-\mu N T}{2}\right)} \left[1-\exp{\left(\frac{-Y_\mathrm{phys}^2}{2 \sigma_D^2}\right)}\right]^{\frac{1}{2}}
\end{equation}

\section{Derivation of imaging geometries}
In an imaging configuration, a ray originating from the sample plane at $(y_s,\alpha_s)$ travels through the objective to the detector plane at $(y_d,\alpha_d)$. This transformation between the two planes can be expressed as:
\begin{equation} \label{eq32}
	\begin{bmatrix}
		y_d \\
		\alpha_d
	\end{bmatrix}
	= 
	\begin{bmatrix}
		1 & d_2 \\
		0 & 1
	\end{bmatrix}
	\mathbf{M^N}
	\begin{bmatrix}
		1 & d_1 \\
		0 & 1
	\end{bmatrix}
	\begin{bmatrix}
		y_s \\
		\alpha_s
	\end{bmatrix}
	\equiv 
	\mathbf{K}
	\begin{bmatrix}
		y_s \\
		\alpha_s
	\end{bmatrix}
\end{equation}

Here, $d_1$ is the distance from the sample plane to the entry of the objective, $d_2$ is the distance from the exit of the objective to the detector plane and $\mathbf{M^N}$ is the RTM for the CRL.

$\mathbf{K}$ is therefore given in terms of $\mathbf{M^N}$ as:
\begin{equation} \label{eq33}
	\mathbf{K} = 
	\begin{bmatrix}
		M_{11}^N + d_2 M_{21}^N & M_{12}^N + d_1 (M_{11}^N+d_2 M_{21}^N) + d_2 M_{22}^N \\
		M_{21}^N & d_1 M_{21}^N + M_{22}^N
	\end{bmatrix}
\end{equation}

The imaging condition implies that:
\begin{equation} \label{eq34}
	\begin{bmatrix}
		0 \\
		\alpha_d
	\end{bmatrix}
	= 
	\mathbf{K}
	\begin{bmatrix}
		0 \\
		\alpha_s
	\end{bmatrix}
\end{equation}	

i.e.:
\begin{equation} \label{eq35}
	K_{12} = 0
\end{equation}

Inserting this leads to:
\begin{equation} \label{eq36}
	f\varphi \sin(N\varphi) + d_1 \left(\cos(N \varphi) - d_2 \frac{\sin(N\varphi)}{f \varphi}   \right) + d_2 \cos (N \varphi)  =  0
\end{equation}

This can rewritten as the \emph{imaging condition} for a thick CRL:
\begin{equation} \label{eq37}
	 \frac{1}{d_1}   +  \frac{1}{d_2} -  \frac{1}{f_N} +  \frac{f \varphi \tan(N \varphi)}{d_1 d_2} = 0
\end{equation}

Next, the magnification of the imaging system $\mathcal{M}$ (a positive number) is defined by:
\begin{equation} \label{eq38}
	K_{11} = \mathcal{M} = -\frac{y_d}{y_s}
\end{equation}

We therefore have a set of two equations:
\begin{equation} \label{eq39}
	\begin{split}
		0 &= M_{12}^N + d_1 (M_{11}^N+d_2 M_{21}^N) + d_2 M_{22}^N \\
		-\mathcal{M} &= M_{11}^N + d_2 M_{21}^N
	\end{split}
\end{equation}

which can then be solved simultaneously to find expressions for $d_1$ and $d_2$:
\begin{equation} \label{eq40}
	d_1 = - \frac{M_{12}^N}{\mathcal{M}} +\frac{M_{11}^N M_{22}^N}{M_{21}^N \mathcal{M}} + \frac{M_{22}^N}{M_{21}^N} = \frac{f \sin{(\varphi)}}{\sin{(N \varphi)}} \left[\cos{(N \varphi)} +\frac{1}{\mathcal{M}}\right] =  f_N \left[1 + \frac{1}{ \mathcal{M}\cos (N \varphi)}\right]
\end{equation}

\begin{equation} \label{eq41}
	d_2 = -\frac{\mathcal{M}-M_{11}^N}{M_{21}^N}  = \frac{f \sin{(\varphi)}}{\sin{(N \varphi)}} [\cos{(N \varphi)} + \mathcal{M}] = f_N \left[1 + \frac{\mathcal{M}}{ \cos (N \varphi)}\right]
\end{equation}

Note that Eqs. \ref{eq40} and  \ref{eq41} are identical for $\mathcal{M} = 1$. Moreover, $d_1$ and $d_2$ may never be smaller than $f_N$. For practical applications, the total sample-to-detector distance: $L = d_1 + d_2 + NT$ may be a fixed parameter, implying that we may want to rewrite the imaging condition as an expression in $L$ and $\mathcal{M}$. Addition of these two equations then gives: 
 \begin{equation} \label{eq42}
 	\frac{L-NT}{f \varphi} \sin(N\varphi) -2\cos(N \varphi) = \mathcal{M} + \frac{1}{\mathcal{M} } 
\end{equation}

We can rewrite this using trigonometric identities (see Appendix):
\begin{equation} \label{eq43}
 	\sqrt{4 + \left( \frac{L-NT}{f \varphi} \right)^2}   \sin \left( N \varphi  - \tan^{-1}( \frac{2f \varphi}{L - NT} ) \right) = \mathcal{M} + \frac{1}{\mathcal{M} } 
\end{equation}

Hence, for a given lenslet (i.e. $f$ and  $\varphi $), total distance $L$ and magnification $\mathcal{M} $, we can determine the required number of lenses from:
\begin{equation} \label{eq44}
 	N =  \frac{1}{\varphi}\left( \sin^{-1}\left(    \frac{    \mathcal{M} + \frac{1} {\mathcal{M}}   }{  \sqrt{( \frac{L-NT}{f \varphi} )^2 +4 }   } \right)    + \tan^{-1}\left( \frac{2f \varphi}{L-NT}  \right)  \right)
\end{equation}

Due to the factor of $N$ in the denominator, $N$ should be determined by evaluating the equation twice \emph{via}, e.g. the fixed-point method. 

\section{Derivation of the acceptance functions for imaging systems}
The expression for $\mathbf{K}$ (Eq. \ref{eq33}) provides the following expression for $y_n$ in terms of the sample plane coordinates $(y_s,\alpha_s)$:
\begin{equation} \label{eq45}
	\begin{split}
		y_n &=  \alpha_s \left[d_1 \left(M^n_{11}-\frac{T}{2}M^n_{21}\right) + M^n_{12}-\frac{T}{2}M^n_{22}\right] + y_s \left(M^n_{11}-\frac{T}{2}M^n_{21}\right) \\
		&= (d_1\alpha_s + y_s)\cos{\left[\left(n-\frac{1}{2}\right)\varphi\right]} + f\varphi \alpha_s \sin{\left[\left(n-\frac{1}{2}\right)\varphi\right]}
	\end{split}
\end{equation}

The cumulative attenuation for the entire imaging system, $U_N$, can be derived in the same manner as for the CRL alone. Temporarily neglecting the physical aperture of the lenslet, this is can be written as: 

\begin{equation} \label{eq46}
	U_N(y_s,\alpha_s) = \exp{(-N \mu T_\mathrm{web})} \times \exp{[-(A_N^* \alpha_s^2 + B_N^* y_s \alpha_s + C_N^* y_s^2)]}
\end{equation}
where the coefficients $A_N^*$, $B_N^*$ and $C_N^*$ are geometrical sums which, using the identities in the Appendix, can be derived as follows:

\begin{equation} \label{eq47}
	\begin{split}
		A_N^* 
		&= \frac{\mu}{R}\sum_{n=1}^N \left[d_1\left(M^n_{11}-\frac{T}{2}M^n_{21}\right)+\left(M^n_{12}-\frac{T}{2}M^n_{22}\right)\right]^2 \\
		&= \frac{\mu}{R}\sum_{n=1}^N \left[d_1(\cos{(n\varphi)}+\frac{\varphi}{2}\sin{(n\varphi)})+f\varphi\sin{(n\varphi)}-\frac{T}{2}\cos{(n\varphi)}\right]^2 \\
		&= \frac{\mu}{R} \left[d_1^2 + (f\varphi)^2\right] \sum_{n=1}^N \cos^2{\left(\left(n-\frac{1}{2}\right)\varphi-\cot^{-1}{\left(\frac{d_1}{f\varphi}\right)}\right)} \\
		&= \frac{\mu}{2R} \left[d_1^2 + (f\varphi)^2\right] \left[N-1+\frac{1}{\varphi}\sin{\left((N+1)\varphi\right)}\cos{\left((N-1)\varphi-2\cot^{-1}{\left(\frac{d_1}{f\varphi}\right)}\right)}\right]
	\end{split}
\end{equation}

\begin{equation} \label{eq48}
	\begin{split}
		B_N^* 
		&= \frac{2\mu}{R}\sum_{n=1}^N\left(M^n_{11}-\frac{T}{2}M^n_{21}\right)\left[d_1\left(M^n_{11}-\frac{T}{2}M^n_{21}\right)+\left(M^n_{12}-\frac{T}{2}M^n_{22}\right)\right] \\
		& = \frac{2\mu}{R}\sum_{n=1}^{N}\left[d_1(\cos(n\varphi)+\frac{\varphi}{2}\sin(n\varphi))+f\varphi\sin(n\varphi)-\frac{T}{2}\cos(n\varphi)\right]\left(\cos(n\varphi)+\frac{\varphi}{2}\sin(n\varphi)\right)    \\
		&= \frac{2\mu}{R}\left[\sum_{n=1}^N d_1 \cos{\left(\left(n-\frac{1}{2}\right)\varphi\right)} + \sum_{n=1}^N f\varphi \sin{\left(\left(n-\frac{1}{2}\right)\varphi\right)} \cos{\left(\left(n-\frac{1}{2}\right)\varphi\right)} \right] \\
		&= \frac{2\mu}{R}\left[\frac{d_1}{2}(N-1) + \sqrt{d_1^2 + (f \varphi)^2} \frac{\sin{((N+1)\varphi)}}{2\varphi}\cos{\left((N-1)\varphi-\cot^{-1}{\left(\frac{d_1}{f\varphi}\right)}\right)}\right]
	\end{split}
\end{equation}

\begin{equation} \label{eq49}
	\begin{split}
		C_N^* 
		&= \frac{\mu}{R}\sum_{n=1}^N \left(M^n_{11}-\frac{T}{2}M^n_{21}\right)^2 = C_N\\
		&= \frac{N\mu}{2R} \left( 1 +  \text{sinc}(2N\varphi) \right)
	\end{split}
\end{equation}

The function $U_N(y_s,\alpha_s)$ is also a tilted 2D Gaussian. It can be described as the product of a constant term $\exp{(-N \mu T_\mathrm{web})}$ and two 1D Gaussian functions: 
\begin{equation} \label{eq50}
	U_N(y_s,\alpha_s) = \exp{(-N \mu T_\mathrm{web})} \times \exp{\left(\frac{-(\alpha_s-\gamma y_s)^2}{2 \sigma_a^2}\right)} \times \exp{\left(\frac{-y_s^2}{2 \sigma_v^2}\right)} 
\end{equation}

These two functions are the \emph{vignetting function}, which describes the reduction in brightness from the centre of the optical axis towards its periphery in terms of an RMS $\sigma_v$, and the \emph{angular acceptance function}, which describes the angle over which the lens collects radiation emitted from a point $y_s$ on the sample plane. The angular acceptance function has RMS $\sigma_a$ and offset $\gamma$. Their individual expressions are derived from the coefficients $A_N^*$, $B_N^*$ and $C_N^*$ in the following subsections.

\subsection{Angular acceptance}
By inspection of Eq. \ref{eq46}, the angular acceptance function is a Gaussian with standard deviation $\sigma_a$ given by:
\begin{equation} \label{eq51}
	\begin{split}
		\sigma_a 
		&= \sqrt{\frac{1}{2 A_N^*} } \\
		&= \sqrt{\frac{R}{\mu N[d_1^2+(f\varphi)^2]}} \left[1+\frac{1}{N}-\frac{1}{N\varphi}\sin{((N+1)\varphi)}\cos{\left((N-1)\varphi+2\tan^{-1}{\left(\frac{d_1}{f\varphi}\right)}\right)}\right]^{-\frac{1}{2}}
	\end{split}
\end{equation}

The last term is a slowly varying function that asymptotically approaches 1 for $N\varphi \to 0$. The analytical expression for $\gamma$ can similarly be expressed as:
\begin{equation} \label{eq52}
	\begin{split}
		\gamma 
		&= \frac{B_N^*}{2A_N^*} \\
		&= \frac{(f+2d_1N)\sqrt{d_1^2+(f\varphi)^2}\cos{\left(2N\varphi+\tan^{-1}{\left(\frac{d_1}{f\varphi}\right)}\right)}}{2[d_1f + N(d_1^2+f^2)]+\frac{1}{\varphi}[d_1^2+(f\varphi)^2]\sin{\left(2N\varphi\right)}}
	\end{split}
\end{equation}

\subsection{Vignetting}
By inspection of Eq. \ref{eq46}, the RMS value $\sigma_v$ is given by:
\begin{equation} \label{eq53}
	\begin{split}
	\sigma_v 
		&= \sqrt{\frac{2A_N^* }{4 A_N^* C_N^*  -  B_N^{*2} }} \\
		&= \frac{\delta}{\mu \sigma_a}\left[(N\varphi)^2-\sin^2{(N\varphi)}\right]^{-\frac{1}{2}}
	\end{split}
\end{equation}

\section{Derivation of spatial resolution}
The spatial resolution, $\Delta y_d$, is the minimum distance two objects can be separated while still being `distinguishable' by the imaging system. While the Rayleigh criterion \cite{born99} provides a simple and effective approximation for this, it is not applicable to systems with Gaussian pupil (i.e. attenuation) functions. Instead, we must first derive the point-spread function (PSF) of the system, then determine the minimum resolvable distance between two such PSFs.

In an ideal imaging configuration, the PSF is simply the Fourier transform of the pupil function, $P(y_0)$ \cite{born99}:
\begin{equation} \label{eq54}
	PSF(y_s) = \mathcal{F}(P(y_0))
\end{equation}

where the pupil function is the acceptance of a ray emanating from the centre of the field-of-view, derived from Eq. \ref{eq50}:
\begin{equation} \label{eq55}
	P(y_0) = \exp{\left(\frac{-y_0^2}{2 \sigma_a^2 d_1^2}\right)} \times H\left(\frac{y_0}{Y_\mathrm{pup}}\right)
\end{equation}

Here, $Y_\mathrm{pup} \equiv Y_\mathrm{phys}/\sqrt{1+\left(f\varphi/d_1\right)^2}$. For a wavenumber $k=2\pi/\lambda$ at x-ray wavelength $\lambda$, and ignoring constant prefactors, the point-spread intensity function at the source plane, $PSF(y_s)$, is then given by:
\begin{equation} \label{eq56}
	\begin{split}
		PSF(y_s) &= \left\lvert \int_{-\infty}^{\infty}{P(y_0)\exp{\left[-i\frac{k}{d_1}y_0 y_s\right]}}\mathrm{d}y_0\right\rvert^2 \\
		&= \exp{\left(-k^2\sigma_a^2 y_s^2\right)} \times \left[\erf{\left(\frac{Y_\mathrm{pup}+i k\sigma_a^2y_s}{\sqrt{2}\sigma_a}\right)}+\erf{\left(\frac{Y_\mathrm{pup}-i k\sigma_a^2y_s}{\sqrt{2}\sigma_a}\right)}\right]^2
	\end{split}
\end{equation}

The resolution can then be defined by the the separation distance between two PSFs corresponding to a contrast ratio of $C$ (where $C$ is small when the contrast is poor). Using Eq. \ref{eq56}, this can be determined by numerically solving:
\begin{equation} \label{eq57}
	\frac{PSF(0)+PSF(\Delta y_s)}{PSF(\frac{\Delta y_s}{2})+PSF(-\frac{\Delta y_s}{2})} = 1-C
\end{equation}

In the case of absorption-limited (i.e. Gaussian) CRLs, this gives a function in terms of $\lambda$, $\sigma_a$ and $C$:

\begin{equation} \label{eq58}
	\Delta y_d = \sqrt{0.06905 -0.1019\log{(1-C)}}\frac{\lambda}{\sigma_a}
\end{equation}

\section{Derivation of chromatic aberration}

The refractive decrement $\delta$ is approximately inversely proportional to the square of the of the x-ray energy. Hence, compared to $\delta_0$ at a nominal energy $E_0$, the actual $\delta$ at energy $E$ will be:
\begin{equation} \label{eq59}
	\delta \approx \delta_0\frac{E_0^2}{E^2} = \frac{\delta_0}{(1+\epsilon)^2}
\end{equation}

Where $\epsilon$ is a (small) normalised energy perturbation such that $E/E_0 = 1 + \epsilon$. This energy dependence of $\delta$ means that the values for $f$ and $\varphi$ in the formalism are similarly chromatic. Their new values values are:
\begin{equation} \label{eq60}
	\varphi = \frac{\varphi_0}{1+\epsilon} \approx \varphi_0(1-\epsilon), \qquad f = f_0(1+\epsilon)^2 \approx f_0(1+2\epsilon)
\end{equation}

A ray originating from the sample at $y_s = 0$ will strike the detector at $y_d$ according to:
\begin{equation} \label{eq61}
	y_d = \alpha_s\left[(d_1+d_2)\cos{(N\varphi)}+\left(f\varphi - \frac{d_1 d_2}{f\varphi}\right)\sin{(N\varphi)}\right]
\end{equation}

If we are in an ideal imaging condition at $E = E_0$, then $y_{d,0} = y_s = 0$. However, if $E \neq E_0$ then the ray will strike the detector at $y_d \neq y_s = 0$. The dependence of $y_d$ on $\epsilon$ is found by substituting the chromatic expressions for $\phi$ and $f$ and Taylor expanding to first order in $\epsilon$:
\begin{equation} \label{eq62}
	y_d = y_{d,0} + d_\mathrm{ch}\epsilon\alpha_s = d_\mathrm{ch}\epsilon\alpha_s
\end{equation}

where:
\begin{equation} \label{eq63}
	d_\mathrm{ch} = \frac{N\varphi_0[d_1 d_2 - (f_0\varphi_0)^2]\cos{(N\varphi_0)}+[d_1 d_2 + (f_0\varphi_0)^2 + f_0N\varphi_0^2(d_1+d_2)]\sin{(N\varphi_0)}}{f_0\varphi_0}
\end{equation}

To calculate the intensity of a chromatic ray, we first consider that the incident ray will have an intensity at the sample $I_\epsilon$ defined by the chromatic distribution of the source. In many cases this is Gaussian with an RMS bandwidth $\sigma_e$:
\begin{equation} \label{eq64}
	I_\epsilon = I_0 \frac{1}{\sqrt{2\pi}\sigma_e}\exp{\left(\frac{-\epsilon^2}{2 \sigma_e^2}\right)}
\end{equation}

The ray will then be attenuated by the material in the lens as a function of its incident angle $\alpha_s$. From the formalism we know that this is a Gaussian attenuation function with RMS width $\sigma_a$ given by:
\begin{equation} \label{eq65}
	U(\alpha_s) = \exp{\left(\frac{-\alpha_s^2}{2 \sigma_a^2}\right)}
\end{equation}

Thus, the intensity of the ray at the detector position, $I_d$, will be:
\begin{equation} \label{eq66}
I_d(\epsilon,\alpha_s) = I_\epsilon \times U(\alpha_s) = I_0 \frac{1}{\sqrt{2\pi}\sigma_e}\exp{\left(\frac{-\epsilon^2}{2 \sigma_e^2}\right)} \times \exp{\left(\frac{-\alpha_s^2}{2 \sigma_a^2}\right)}
\end{equation}

From Eq. \ref{eq61}, we can calculate the initial angle of a ray given its $\epsilon$ and $y_d$ as $\alpha_s = y_d/(d_\mathrm{ch}\epsilon)$. Substituting this into Eq. \ref{eq66} then gives the spatio-chromatic intensity distribution at the detector:
\begin{equation} \label{eq67}
	I_d(\epsilon,y_d) = I_0 \frac{1}{\sqrt{2\pi}\sigma_e} \exp{\left(\frac{-\epsilon^2}{2 \sigma_e^2}\right)} \times \exp{\left[\frac{-y_d^2}{2(\sigma_a d_{ch} \epsilon)^2}\right]}
\end{equation}

Normalising $y_d$ by the system magnification $\mathcal{M}$ then gives the perceived chromatic point-spread in the sample coordinates, $y_s$:
\begin{equation} \label{eq68}
	I_s(\epsilon,y_d) = I_0 \frac{1}{\sqrt{2\pi}\sigma_e} \exp{\left(\frac{-\epsilon^2}{2 \sigma_e^2}\right)} \times \exp{\left[\frac{-y_s^2}{2(\sigma_a d_{ch} \epsilon \mathcal{M})^2}\right]}
\end{equation}

So the chromatic point-spread function ($PSF_\mathrm{chr}$) is then obtained by integrating $I_d$ across $\epsilon$:
\begin{equation} \label{eq69}
	\begin{split}
		PSF_\mathrm{ch} &= I_0 \frac{1}{\sqrt{2\pi}\sigma_e} \int_{-\infty}^{\infty}{\exp{\left(\frac{-\epsilon^2}{2\sigma_e^2}\right)} \times \exp{\left(\frac{-y_s^{2}}{2(\sigma_a d_\mathrm{ch}\mathcal{M} \epsilon)^2}\right)}\mathrm{d}\epsilon} \\
		&= I_0 \exp{\left(\frac{-\lvert y_s \rvert}{\sigma_e \sigma_a \mathcal{M} d_\mathrm{ch}}\right)}
	\end{split}
\end{equation}

Which is a Laplacian distribution with characteristic width:
\begin{equation} \label{eq70}
	\sigma_\mathrm{ch} = \sigma_a \sigma_e \mathcal{M} d_\mathrm{ch}
\end{equation}

\section{Conclusions}

In conclusion, this formalism describes the focusing and attenuation properties of CRLs and CRL-based imaging systems. Specifically, it provides closed analytical expressions for crucial optical parameters including vignetting, angular acceptance and both chromatic and diffraction-limited spatial resolution.

The expressions provided are general and may be used on the vast majority of imaging configurations, including at the `thick-lens limit', where working distances are very small compared to the length of the CRL.

Moreover, these expressions are highly suited for fast computation as well as algebraic manipulation, and may therefore be used for analytical and parametric optimisation.

\appendix*
\section{Trigonometric Identities}

Throughout this work, we will make use of the following identities:
\begin{equation}\label{eq1a}
	a \cos(\theta) + b \sin(\theta) = \sqrt{a^2 + b^2} \cos\left(\theta - \cot^{-1}\left(\frac{a}{b}\right)\right)
			= \sqrt{a^2+b^2} \sin \left(\theta + \tan^{-1}\left(\frac{a}{b}\right)\right)
\end{equation}

And:
\begin{eqnarray} \label{eq2a}
	\sum_{n=1}^N \sin^2(n\theta + d) = \frac12 \left(N+1 - \frac{\sin((N+1)\theta)}{\sin\theta} \cos(N\theta+2d) 				\right)\\
	\sum_{n=1}^N \cos^2(n\theta + d) = \frac12 \left(N-1 + \frac{\sin((N+1)\theta)}{\sin\theta} \cos(N\theta+2d) 				\right)\\
	\sum_{n=1}^N \cos(n\theta + d) \sin(n\theta + d) = \frac12 \left( \frac{\sin((N+1)\theta)}{\sin\theta}
			\sin(N\theta+2d) \right)	
\end{eqnarray}

Furthermore, we utilize that $\tan^{-1}(x) = \cot^{-1}\left(\frac{1}{x}\right) \approx x$ for $x \ll 1$ 

Finally we mention the following two identities

\begin{eqnarray} \label{eq3a}
\sin( \cot^{-1}(x) ) = \frac{1}{\sqrt{1 + x^2}} \\
\cos( \cot^{-1}(x) ) = \frac{x}{\sqrt{1 + x^2}} \\
\end{eqnarray}

\begin{acknowledgments}
We thank Frederik St\"{o}hr and Ray Barrett for useful discussions. In addition, we are grateful to the ESRF for providing beamtime on ID06 and Danscatt for travel funding. H.F.P and S.R.A. acknowledge support from the E.R.C. grant d-TXM'. H.S. acknolwedges support from a DFF-FTP individual postdoc grant.
\end{acknowledgments}

\bibliography{article_HS}

\end{document}